\RequirePackage{lineno} 
\documentclass[twocolumn,showpacs,preprintnumbers,amsmath,amssymb]{revtex4}

\usepackage{graphicx}
\usepackage{dcolumn}
\usepackage{bm}
\usepackage[usenames]{color}

\begin{document}



\title{Searching for doubly-charged vector bileptons

in the Golden Channel at the LHC

}

\author{B. Meirose}
\email{Bernhard.Meirose@cern.ch}
\affiliation{
Department of Physics \\
Division of Experimental High-Energy Physics \\
S\"olvegatan 14, 223 62 Lund, Sweden
}%

\author{A. A. Nepomuceno}%
\email{Andre.Asevedo@cern.ch}


\affiliation{%
INFES \\
Universidade Federal Fluminense \\
Cx.Postal 68528, Santo Antonio de Padua, RJ, Brazil
}%

\date{\today}

\begin{abstract}
In this paper we investigate the LHC potential for discovering doubly-charged vector bileptons considering the measurable 
process $p,p$ $\rightarrow$ $e^{\mp}e^{\mp}\mu^{\pm}\mu^{\pm} X$. We perform the study using four different bilepton masses 
and three different exotics quark masses. Minimal LHC integrated luminosities needed for discovering and for setting limits 
on bilepton masses are obtained for both 7 TeV and 14 TeV center-of-mass energies. We find that these spectacular signatures 
can be observed at the LHC in the next years up to a bilepton mass of order of 1 TeV.
\end{abstract}


\maketitle
\section{\label{sec:level1}Introduction:\protect\\
} 

Although bileptons \cite{CUYDAV} can be considered as “exotic” particles relative to their Standard Model (SM) cousins due to 
their unfamiliar quantum numbers, they are, in a sense, a conservative prediction. Even though they are suggested only by 
special extensions of the SM, their existence is, in our view, no more special than other proposals such as weak-scale extra 
dimensions or supersymmetry (SUSY). Indeed bileptons are a prediction employing only the model building rules for 
renormalizable gauge theories so successful in the Standard Model. The main motivation for expecting bileptons 
is that they explain three quark-lepton families. 

Generically speaking, a bilepton is a boson which couple to two leptons, but not to SM quarks, and which carries two 
units of lepton number. They are present in several beyond-SM scenarios, such as left-right symmetric models, technicolor 
and theories of grand unification. The bileptons in which we are interested are doubly charged vector bosons which couple 
to SM leptons, and are predicted when the Standard Model is embedded in a larger gauge group. The so-called 331 
models \cite{PIPLEI, FRA} fall into this category and in this article we restrict ourselves to this case. However, 
as will be explained in section II, we expect our results to hold in any model containing vector bileptons.

In bilepton pair production, each of the bileptons will decay to two same-sign leptons. 
Therefore, they provide an exceptionally clean signature of four isolated high transverse momentum leptons, not necessarily of the same flavor.
In this article we explore some of the consequences of this fact at the LHC. We study the actual collider signatures for the process 
$p,p$ $\rightarrow$ $e^{\mp}e^{\mp}\mu^{\pm}\mu^{\pm} X$ which has no SM background and is for this reason a ``golden'' 
channel for finding bileptons. In this process, a bilepton pair is produced via $s$ or $t$ channel where one of them decays into electrons while the other decays into muons. 

The most useful current lower bound on vector bileptons require these particles to be heavier than 740 GeV \cite{TUL}. 
This limit has been derived from experimental limits on fermion pair production at LEP and lepton-flavor charged lepton 
decays. Another useful lower bound is $M_Y >$ 850 GeV, a result which was established from muonium-antimuonium 
conversion \cite{WILL}. Although more stringent, this limit depends on the assumption that the bilepton coupling is 
flavor-diagonal. In this article we nevertheless consider bilepton masses as low as 400 GeV, following a similar line of 
reasoning as \cite{YARA}, where the authors argued a lower bound of 350 GeV for doubly-charged vector bilepton masses, 
a limit that is compatible with other low energy bounds \cite{LOWE}. We also allow a larger upper bound for bilepton 
masses in 331 models than the usual 1 TeV considered by some authors.

This article has been organized as follows. In section II we present our motivations to perform the present study. 
In section III we explain the numerical procedure for simulating the $p,p$ $\rightarrow$ $e^{\mp}e^{\mp}\mu^{\pm}\mu^{\pm} X$ 
reaction as well as its validation. In section IV we show relevant experimental observables for the bilepton golden channel. 
In section V we present the discovery potential for bileptons, calculating mass exclusion limits as a function of the LHC 
integrated luminosities, including a digression on the prospects for the accelerator's 7 TeV run and for 
the super LHC (sLHC). We conclude in section VI.

\section{Motivations}
\par
Many interesting channels have been studied in the literature concerning bileptons, but curiously, 
there has been no systematic study on the phenomenology of the bilepton golden channel at the LHC. 
The authors in \cite{DION} did a fairly comprehensive study of bilepton phenomenology at hadron colliders. 
But in all cases they limited themselves to bilepton pair-production study, disregarding its decays. 
In the present study we go a step further in understanding the actual collider signatures, 
by considering measurable final states. 

\subsection{331 models}

The 331 models are based on the gauge symmetry $SU(3)_C \otimes SU(3)_L \otimes U(1)_X$, hence their name. Their first versions appeared in 1992 \cite{PIPLEI, FRA}. There are many interesting aspects of the 331 models worth noticing but the most intriguing one is the explanation of three quark-lepton families, which is the main motivation for expecting bileptons in Nature. This is done via a nontrivial anomaly cancellation in the 331 Model that takes place between families, which is achieved by requiring the number of families to be equal to the number of quark colors. The explanation of the number of generations is also arguably one of the main reasons that keep model builders interested in 331 models, since they are one of the few which elegantly address this problem. The 331 models are also the simplest extension of the SM containing bileptons. Other interesting features of the 331 models include: a) they treat the third generation differently than the first two, this lead to an explanation of the heavy top quark mass; b) they have an automatic Peccei-Quinn symmetry \cite{PQ}, hence they are also able to solve the strong CP problem. Moreover, gauge symmetry $SU(3)_C \otimes SU(3)_L \otimes U(1)_X$ is considered a subgroup of the popular $E_6$ \cite{E6} Grand Unified Theory (GUT), which can be itself derived from $E_8 \otimes E_8$ \cite{E8} heterotic string theory. Finally, it is worth mentioning that contrary to the SM, in 331 models lepton family number is not required to be conserved, only total lepton number. There is already experimental proof that lepton family number is not an exact symmetry via neutrino oscillations and one can regard this as circumstantial evidence for the non conservation of lepton family number in general. The combination of such intriguing aspects make bileptons desirable candidates to be found in Nature.

Another important point to be discussed is the minimal version of the 331 models. There are different ways in which $SU(3)_C \otimes SU(3)_L \otimes U(1)_X$ can be broken down back to the $SU(3)_C \otimes SU(2)_L \otimes U(1)_Y$ SM gauge symmetry. The minimal version corresponds to use minimal Higgs structure to achieve this goal. In this version it is required that the new neutral vector boson $Z^\prime$ mass term to be coupled to the bilepton mass, like:

\begin{equation}
{M_Y \over M_{Z^\prime}} = {{\sqrt{3(1-4\sin^2\theta_W)} \over
2\cos \theta_W }}
\end{equation}

Regarding theoretical upper bounds in 331 models there is no consensus in the literature. 
It was reasoned in \cite{FRA} that in 331 models, bileptons cannot be significantly heavier than 1 TeV, 
because of an upper limit in the symmetry breaking scale which is placed by requiring the sine squared of the 
Weinberg mixing angle ($\theta_W$) to be smaller than 1/4 , which is the same line of argumentation used in \cite{NG} 
to conclude that the $Z^\prime$ mass cannot be itself heavier than 3.1 TeV. Considering the mass relation 
between $Z^\prime$ and the bilepton, it could be argued that at least the minimal version of the 331 model could be 
excluded, should bileptons not be detected at the LHC, since any vector bilepton mass heavier than $\sim$ 840 GeV 
would violate the $Z^\prime$ mass upper limit via the mass relation between the two gauge bosons given
by equation (1).
This conclusion was challenged in \cite{PLEITEZ}. The argument is as follows. The 331 model predicts that there is an 
energy scale $\mu$ where the model loses its perturbative character. Should experimental data suggest a lower bound on 
the vector bilepton mass larger than $\mu$, the model would be ruled out. The value of $\mu$ is calculated through 
the condition $\sin^2\theta_W(\mu)=1/4$, but from this requirement alone it is not possible to know the real value 
of $\mu$. Then the upper limit on the vector bilepton mass could be, for instance, 3.5 TeV, as has 
been discussed by \cite{JAIN}. By the same token the 3.1  TeV upper limit in the $Z^\prime$ mass is automatically challenged. 
Therefore we do not believe it is possible to unambiguously discard \textit{any} 331 model at the LHC 
(although they could be \textit{discovered} at it), since we consider the bilepton mass upper bound to reasonably lie 
beyond the accelerator's reach.

For the lower bounds on vector bileptons we consider at least two mass points that violate the general 740 GeV 
limit imposed by LEP data. As explained by the authors in \cite{YARA}, all the constraints on the 331 parameters coming 
from experiments involving leptonic interaction should be examined with caution. In the 331 model the leptons mix by a 
Cabibbo-Kobayashi-Maskawa-like mixing matrix whose elements have not yet been measured, so usually these experiments 
(and derived limits) apply only when the leptonic mixing matrix is diagonal. Also, in models with an extended Higgs sector 
some not unrealistic situations could exist in which the scalar bosons contribution to muonium to antimuonium conversion is 
not negligible. This puts also the possibility of strengthening bilepton experimental limits 
still at the LHC's 7 TeV run in a new perspective, a possibility that we also discuss in this article.

\subsection{331 Models and Supersymmetry}

In recent years, a considerable fraction of both the experimental and theoretical communities has dedicated itself to supersymmetry. It is doubtlessly the mainstream subject in particle physics. The 331 models are not \textit{necessarily} supersymmetric. But any renormalizable gauge theory can be extended to a globally supersymmetric model. The 331 models, being anomaly free, are renormalizable and fall of course in this category. Some authors have explored this possibility \cite{MSUSY331}. Furthermore, as it was argued also in \cite{PLEITEZ}, in this model, the “hierarchy problem” is less severe than in the SM and its extensions since no arbitrary mass scale can be introduced. The masses of fundamental scalars are sensitive to the mass of the heaviest particles which couple directly or indirectly with them. Since in the 331 model the heaviest mass scale is of the order of a few TeVs there is not a “hierarchy problem” at all. This feature remains valid when supersymmetry is introduced. Thus, the breaking of the supersymmetry is also naturally at the TeV scale in the 331 model.

\subsection{Model independent vector bilepton searches at the LHC}

Ideally one would like to study bileptons in hadron colliders as model independent as possible. For vector bileptons 
this is not possible as the non-inclusion of the $Z^\prime$ boson makes bilepton pair production to violate unitarity, 
in complete analogy to what happens with $e^+e^- \to W^+W^-$ using only photon exchange. This is another motivation to 
use the 331 model, although we do not restrain ourselves to the minimal version, allowing $M_Y$ and $M_{Z^\prime}$ to 
vary independently of one another. Even though model-dependent, our cross-sections should approximately be in the same 
order of magnitude with any other model containing bileptons, since at hadron colliders these particles have to be 
produced by the same Drell-Yan pair production process, regardless of the model. Exotic heavy quark exchange can 
influence this scenario, a possibility we do explore and which make our conclusions even more general.

\section{Numerical implementation and validation}
\par

To simulate the bilepton golden channel at the LHC we have implemented the 331 model in the Comphep generator \cite{COMPHEP}. 
We followed reference \cite{LONG} to implement the bilepton trilinear gauge interactions and \cite{DION} for 
the $Z^\prime$ couplings with fermions. For the bilepton interaction with leptons we have used the Lagrangian expression 
given in \cite{FRANG}, generating the respective couplings using the Lanhep package \cite{LANHEP}. 
We also take into account bilepton interactions with exotic quarks in the 331 model. 
For this, we considered the following interactions with the exotic quark sector:

\begin{equation}
{\cal L}_Q  =  -\frac{g}{2\sqrt 2}[\overline{Q}^c\gamma^\mu\left(1 - \gamma_5\right)qY^{++}_\mu] + \mbox{H. c.},
\label{ll}
\end{equation}

\noindent
where Q ($D_1$,$D_2$,$T_1$) is an exotic heavy quark in the 331 model, q is a SM quark and $g = e\ / sin{\theta_W}$ 
as in the SM. We considered the respective interaction pairs for ($Q,q$) to be ($D_1$,$u$), ($D_2$,$c$) and ($T_1$,$b$), 
where $u$, $c$ and $b$ are SM quarks. Our purpose in including such interactions was, besides studying the influence of 
the heavy quark sector, to guarantee that all relevant quark subprocesses $q\overline{q} \to Y^{++}Y^{--}$ 
respect unitarity. Since the 331 model itself does not determine the elements of the mixing matrix 
(which determines how bileptons interact with exotic quarks), this is a reasonable criteria. This was needed 
for $u\overline{u} \to Y^{++}Y^{--}$, $b\overline{b} \to Y^{++}Y^{--}$ and $c\overline{c} \to Y^{++}Y^{--}$ which violate 
unitarity otherwise. In Figure 1, one can see the $u\overline{u} \to Y^{++}Y^{--}$ reaction cross-section as a function 
of the center of mass energy. Note that up to energies beyond the LHC designed center-of-mass-energy of 14 TeV, 
the cross-section dependence with energy behaves as expected. We tested this fact for all quarks involved in the proton 
parton distribution function (PDF), CTEQ6l1 \cite{CTEQ6l1} that was used for the complete 
$p,p$ $\to$ $e^{\mp}e^{\mp}\mu^{\pm}\mu^{\pm} X$ reaction simulation.

Concerning particle parameters, we considered heavy quark masses to be equal to 400 GeV, 600 GeV and 800 GeV 
(lower bounds on exotic supersymmetric particles impose a lower bound on 331 exotic quark masses to be $\sim$ 
250 GeV \cite{DAS}) and used 1 TeV for the $Z^\prime$ mass, since 331 $Z^{'}$ masses below 920 GeV were excluded 
using results from the CDF collaboration \cite{GUTI}. For the doubly charged bileptons $Y^{++}$ we considered four mass points: 
400, 600, 800, GeV and 1 TeV. For each mass point, $10000$ $p,p$ $\to$ $e^{\mp}e^{\mp}\mu^{\pm}\mu^{\pm} X$ events were generated with Comphep.

The $Y^{++}$, $Z^\prime$ and exotic quarks widths were calculated directly in Comphep for each mass point.The $Z^\prime$ width for $M_Q$ = 400 GeV is $\varGamma_{Z^\prime} \sim$ 360 GeV.  For the other two exotics quark masses, $\varGamma_{Z^\prime}$ drops to $\sim$ 155 GeV, since Z' decay into exotic quarks becomes kinematically  forbidden. The variation of $\varGamma_{Z^\prime}$ with respect to $M_Y$ is of order of 1\% between the highest and lowest bilepton mass considered. 

To further cross-check our implementation we reproduced the results from reference \cite{DION} Figure 4, for bilepton pair production at the LHC using $M_{Z^\prime}$ = 1 TeV. Minor numerical differences in the cross-sections can easily be explained by the use of different PDF's. 
 
\begin{figure}
\rotatebox{-360}{\scalebox{0.35}{\includegraphics{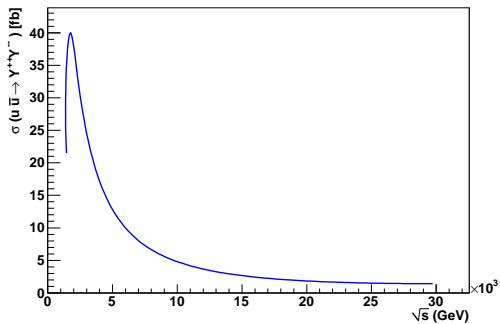}}}
\caption{\label{fig1} $u\overline{u} \to Y^{++}Y^{--}$ cross section as a function of $\sqrt{s}$. The subprocess correctly respects unitarity.}
\end{figure}


\section{Observables}

In what follows, all histograms were produced considering an integrated luminosity of $100\,\textnormal {fb}^{-1}$
and the nominal LHC energy of 14 TeV, unless otherwise stated.

\subsection{Cross section and Width}
\begin{figure}
\rotatebox{-360}{\scalebox{0.35}{\includegraphics{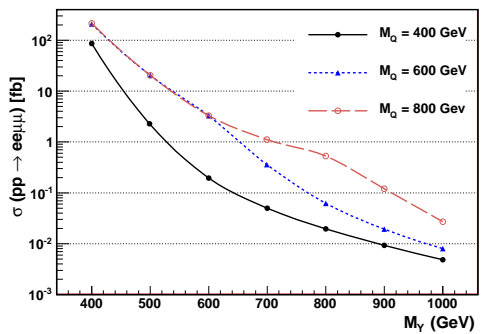}}}
\caption{\label{fig2} Total cross-section of the process $p,p$ $\rightarrow$ $e^{\mp}e^{\mp}\mu^{\pm}\mu^{\pm} X$ as 
a function of bilepton mass. The black/solid, blue/dotted and red/dashed lines represent the cross-section 
for $M_{Q}$ = 400, $M_{Q}$ = 600 and $M_{Q}$ = 800 GeV, respectively.  }
\end{figure}

\begin{figure}
\rotatebox{-360}{\scalebox{0.35}{\includegraphics{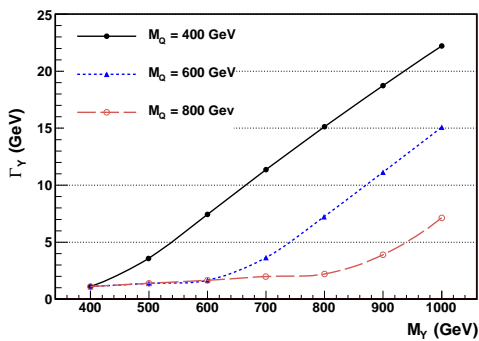}}}
\caption{\label{fig3} Bilepton width as a function of bilepton mass for three different quark mass values.}
\end{figure}

Figure \ref{fig2} shows the total cross-section of the $p,p$ $\rightarrow$ $e^{\mp}e^{\mp}\mu^{\pm}\mu^{\pm} X$ 
process as a function of the doubly-charged bilepton mass, for three different exotic quark masses. 
Here one can see clear evidence on the problem of the influence of the heavy quark sector on bilepton production. 
Note that for a bilepton of $M_Y$ = 800 GeV the effect on the cross-section on having \textit{heavier} 
exotic quark masses of $M_Q$ = 800 GeV as compared to $M_Q$ = 400 GeV is to \textit{increase} the cross-section 
of the the process by a factor of $\sim$ 30. We can also see that the $M_{Q}$ = 600 and $M_{Q}$ = 800 curves split 
at $M_Y$ = 600 GeV. This happens because when $M_{Y} > M_ {Q}$, bilepton decays 
like $Y^{\pm \pm} \rightarrow qQ$ becomes kinematically allowed, which makes the value of 
$Br(Y^{\pm \pm} \rightarrow l^{\pm}l^{\pm}$) decrease for a given bilepton mass.

Figure \ref{fig3} shows the bilepton width as a function of bilepton mass for the same exotic quark masses as before. 
Here we can see how the width increases when new decays are allowed. It is also clear from the plot that the 
bilepton resonance is very narrow.

We admittedly used a very simple approach to the problem of how bileptons and heavy quarks actually mix. However, different ways on determining the values of the mixing matrix between bileptons and heavy quarks could, in principle, intensify such effects even more drastically. An open problem that would deserve a 
separate study of its own.

\subsection{Pseudorapidity}
\begin{figure}
\rotatebox{-360}{\scalebox{0.35}{\includegraphics{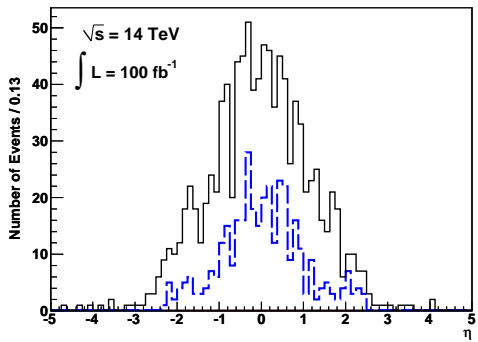}}}
\caption{\label{fig7} Electron pair pseudorapidity distribution for $M_Y$ = $M_{Q}$ = 600 GeV before (black/solid line) 
and after (blue/dashed line) selection . }
\end{figure}

In order to investigate a more realistic scenario, we require  the events that have been generated to pass 
some selection criteria according to the LHC detectors. First, the four leptons must be within  the detectors 
geometrical acceptance, i.e., $\mid \eta \mid < 2.5$ \cite{ATLAS}. With this requirement, the fraction of selected events 
goes from 83\% for $M_Y$ = 400 GeV to 93\% for  $M_Y >$ 900 GeV. Additionally, we also require each lepton $p_T$ be 
greater than 20 GeV. The loss of efficiency due this cut is negligible. Finally, we assume a reconstruction
efficiency of 60\%. Figure \ref{fig7} illustrates the electron pair pseudorapidity distribution of the events before 
and after the selection for $M_Y$ = 600 GeV where we can see the fraction of surviving events. The acceptance 
(geometrical acceptance $\times$ efficiency) for this case is 53\%, and this changes by $\pm$ 3\% 
depending on the bilepton mass. The pseudorapidity distribution for muons is similar. 

\subsection{Transverse Momentum}

\begin{figure}
\rotatebox{-360}{\scalebox{0.35}{\includegraphics{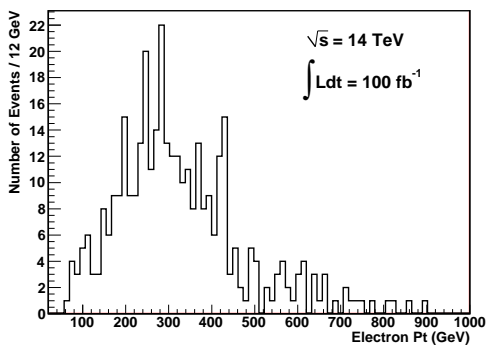}}}
\caption{\label{fig3b} Electron pair transverse momentum distribution for $M_Y$ = $M_{Q}$ = 600 GeV.}
\end{figure}

\begin{figure}
\rotatebox{-360}{\scalebox{0.35}{\includegraphics{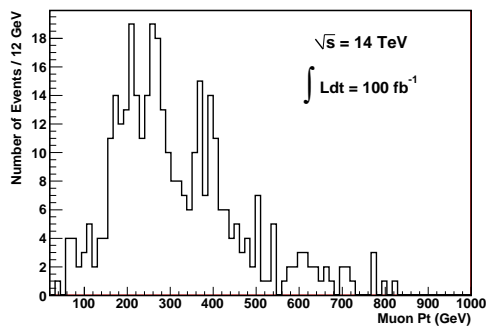}}}
\caption{\label{fig4} Muon pair transverse momentum distribution for $M_Y$ = $M_{Q}$ = 600 GeV.}
\end{figure}
In Figures \ref{fig3b} and \ref{fig4} it is respectively shown the transverse momentum distribution for both the final state selected electrons and muons pairs of the golden channel, where both the doubly-charged bilepton and the heavy quarks have a mass of 600 GeV. Typically most of the events are produced in a region between $\sim$ 200 GeV and 400 GeV but the tail of the distribution has few events going up to 900 GeV for the electron pairs.


\subsection{Invariant Mass}
\begin{figure}
\rotatebox{-360}{\scalebox{0.35}{\includegraphics{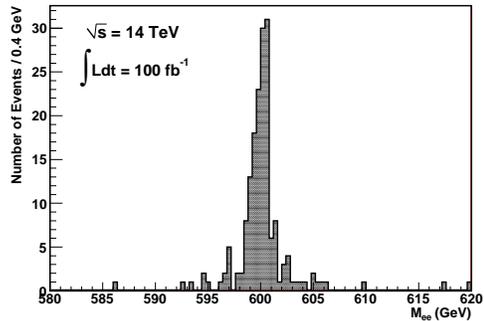}}}
\caption{\label{fig8} Invariant mass for electron pair, for a bilepton of 600 GeV. }
\end{figure}

\begin{figure}
\rotatebox{-360}{\scalebox{0.35}{\includegraphics{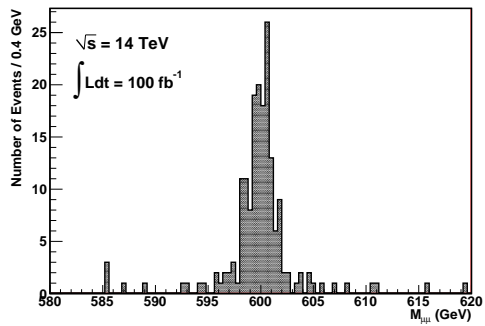}}}
\caption{\label{fig9} Invariant mass for muon pair, for a bilepton of 600 GeV. }
\end{figure}
When doubly-charged vector bileptons are produced in pairs each bilepton can decay to a pair of same-sign leptons, 
not necessarily of the same flavor, where each lepton pair will have the same invariant mass distribution. This would be the most 
compelling evidence of a new resonance coming from a bilepton and very strong evidence of new physics. 
This is displayed in Figures 7 and 8 for the final electron and muon pairs respectively, considering both the bilepton 
and exotic quarks masses to be equal 600  GeV. The mean of the invariant mass for both the electrons and muon pairs in 
the final state unmistakably peaks at 600  GeV, the bilepton mass. Such a plot will of course demand several years of data 
taking  such that enough statistics can be gathered, especially for higher bilepton masses, as we will address in the next 
section.

\section{Discovery Potential and Limits}

\begin{figure}
\rotatebox{-360}{\scalebox{0.35}{\includegraphics{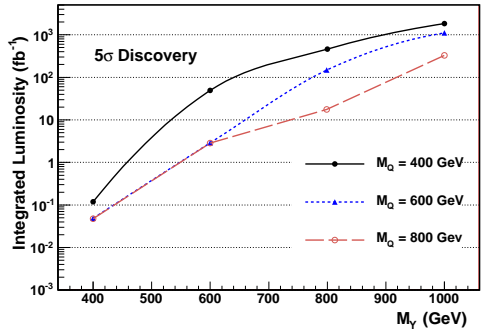}}}
\caption{\label{fig14} Minimal integrated luminosity needed for a $5\sigma$ bilepton discovery at the LHC.}
\end{figure}

In order to determine the LHC potential to find bileptons, we calculate the minimal LHC integrated luminosity needed 
for a five-sigma bilepton discovery. For each bilepton mass, the detector acceptance as stated in section IV.B is considered 
and the $5 \sigma$ significance is obtained by requiring 5 events with two electrons and two muons in the final state
to be produced. The minimal integrated luminosity $L_{int}$ is given by

\begin{equation}
L_{int}  =   \frac{5}{\varepsilon(M_Y) \sigma(M_Y)}
\end{equation}

\noindent
where $\varepsilon(M_Y)$ is the detector acceptance and $\sigma(M_Y)$ is the cross seciton. Figure \ref{fig14} shows the 
calculated values of $L_{int}$ as a function of bilepton mass. From the plot, we  conclude  that a integrated 
luminosity of order of 10 $\textnormal{pb}^{-1}$ is enough for 
discovering a bilepton of 400 GeV mass, which means that such signal can be observed at the very early days of 
LHC running with 14 TeV, even in a regime of low luminosity. Depending on the exotic quark mass, luminosities of  
order 10 $\textnormal{fb}^{-1}$ to 100 $\textnormal{fb}^{-1}$ are needed to discover bileptons if $M_{Y}$ =  800 GeV. 
These scenarios can be achieved after 1 year of LHC operation in low  and high luminosity regimes, respectively.
Finally, for $M_{Y}$ = 1 TeV, around 10 years of LHC operation at high luminosity would be needed for the 
resonance observation. This is in contrast with what would happen at the ILC where M\o ller and 
Bhabha scattering receive both huge corrections  from virtual vector bileptons \cite{MEIROSE}, so that the bilepton 
mass reach can be as high as $\sim$ 11 TeV, provided polarized beams are used.

\begin{figure}
\rotatebox{-360}{\scalebox{0.35}{\includegraphics{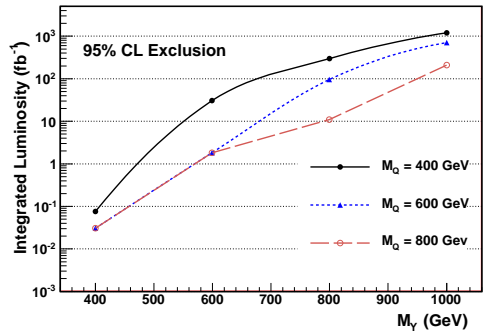}}}
\caption{\label{fig15} Integrated luminosity needed for 95\% CL bileptons exclusion taking into account acceptance and luminosity uncertainties. }
\end{figure}

Figure \ref{fig15} shows the integrated luminosity required  for excluding bileptons at 95\% CL as 
function of the bilepton mass. For this estimation, we have used the D0 limit calculator \cite{D0} to set upper limits 
on the cross sections that are consistent with the observation of zero events in data, assuming no background. 
The Bayesian technique is used to set this limits. In this approach, given a posterior probability density function for 
the signal cross-section, the upper limit on the signal cross section $\sigma_{\textnormal{up}}$, specified at some 
confidence level $100 \times \beta\%$, is given by 

\begin{equation}
\beta = \int_0^{\sigma_{\textnormal{up}}} d\sigma \rho(\sigma \arrowvert k,I)
\end{equation}

\noindent
where $\beta$ = 0.95, $\rho(\sigma \arrowvert k,I)$ is the posterior probability density function, $k$ is the number 
of events observed and $I$ represents prior information available. 

The acceptances values used in this calculation are the same as before.
We have assumed an uncertainty of 5\% on the acceptance and 10\% on the integrated luminosity. 
It also is assumed that the errors on the acceptance and luminosity are uncorrelated. The obtained limits on 
the cross section are then translated to limits on the bilepton mass. Comparing Figure \ref{fig15} with the 
discovery plot, we can see that around 60\% of the discovery integrated luminosity is needed for excluding 
bileptons of a given mass. 

If a significant signal is observed, the next natural step is to determine the properties of 
the new particle. In order to check how well the bilepton mass can be reconstructed with the amount of data needed for 
discovering, we perform a unbinned maximum likelihood fit to the $M_{ee}$ and $M_{\mu \mu}$ distributions 
for $M_{Y}$ = 600 GeV and $M_{Y}$ = 800 GeV MC samples, with a fixed exotic quark mass of 600 GeV. 
The probability density function used in the fits is a Breit-Wigner and two parameters are fitted: the position of the 
invariant mass peak $m$ and the resonance width $\varGamma$. For each bilepton mass, fits are performed to 1000 MC 
experiments (i.e, 1000 $M_{ee}$ and $M_{\mu \mu}$ distributions) and the mean values of the fitted parameters are reported. The number of events in each MC experiment is fixed to 5.
Table I shows the mean, $\bar{m}$, of the fitted mass values and the standard deviation of the $m$ distribution. 
For both bilepton masses, we see that there is a very good agreement between fitted and true masses in both channels. 
As expected, the spread of the distribution is larger for $M_{Y}$ = 800 GeV. The bilepton width can also be 
obtained from the fit at generator level, but it will be dominated by the detector resolution in a 
more realistic scenario, since bileptons are very narrow resonances.

\begin{table}
  \begin{center}
    \caption[]{%
      Mean and standard deviation of fitted mass peak for dielectron and dimuon invariant mass distributions.
    \label{tab:tab1}}
\begin{tabular}{|c|c|c|c|c|}
\hline
&\multicolumn{2}{c|}{$M_{Y}$ = 600 GeV}&\multicolumn{2}{c|}{$M_{Y}$ = 800 GeV}   \\
\hline
channel  & $\bar m$ & $\sigma{_m}$ & $\bar m$ & $\sigma_{m}$   \\ \hline
$ee$     & $599.9$     & $0.5$        &   $799.9$   & $2.4$  	   \\ \hline    
$\mu\mu$ & $600.1$     & $0.6$        &   $799.9$   & $2.6$  	    \\ \hline
   
\end{tabular}
\end{center}
\end{table}

\subsection*{LHC 7 TeV run potential}
Considering the LHC's goals until the end of 2012 for a center-of-mass energy of 7 TeV, we estimate the potential for 
discovering or for setting limits on bilepton masses and couplings at this data taking stage. We consider three scenarios 
with 1, 5 and 10 $\textnormal{fb}^{-1}$ \cite{LHC} of integrated luminosity. Using the D0 limit calculator and using the 
same values for acceptance and uncertainties as in the previous sections, we obtain the respective bilepton masses 
consistent with the observation of zero events for three exotic quark masses: 427, 466 and 483 GeV for $M_{Q}$ = 400 GeV; 478, 534 
and 566 GeV for both $M_{Q}$ = 600 GeV and $M_{Q}$ = 800 GeV, from the lowest to the highest integrated luminosity, respectively. 
These are the doubly-charged vector bilepton masses that can be excluded at 95\% CL. The 7 TeV exclusion limits are 
summarized in Table II.

\begin{table}
  \begin{center}
    \caption[]{%
      Exclusion limits for doubly-charged bileptons with respect to LHC's integrated luminosity and 331 exotic quark masses at the 7 TeV run. Masses are in GeV.
    \label{tab:tab2}}
\begin{tabular}{|c|c|c|c|c|}
\hline
&\multicolumn{3}{c|}{Integrated luminosity} \\
\hline
$M_{Q}$   & 1 $\textnormal{fb}^{-1}$  & 5 $\textnormal{fb}^{-1}$   & 10 $\textnormal{fb}^{-1}$  \\ \hline
$400$              & $427$   &  $466$     &   $483$ 	   \\ \hline    
$600$              & $478$   &  $534$     &   $566$ 	    \\ \hline
$800$              & $478$   &  $534$     &   $566$ 	    \\ \hline   
\end{tabular}
\end{center}
\end{table}

The $5 \sigma$ discovery potentials at 5 and 10 $\textnormal{fb}^{-1}$ of integrated luminosity are respectively found to 
be 452, 535 GeV for $M_{Q}$ = 400 GeV, 511, 542 GeV for $M_{Q}$ = 600 GeV and 515, 544 GeV for $M_{Q}$ = 800 GeV. With 
1 $\textnormal{fb}^{-1}$ the reach is 459 GeV using the highest exotic quark mass. This mass reach is way above the minimum bound of 
350 GeV, so such a discovery at this phase, is not completely discarded. In any case, even if no discoveries are made at 
the 7 TeV run, the exclusion limits that will be established are still valuable for setting up the scenario for the 14 
TeV run. The 7 TeV discovery reach results are summarized in Table III.

\begin{table}
  \begin{center}
    \caption[]{%
      Discovery mass reach for doubly-charged bileptons with respect to LHC's integrated luminosity and 331 exotic quark masses at the 7 TeV run. Masses are in GeV.
    \label{tab:tabX}}
\begin{tabular}{|c|c|c|c|c|}
\hline
&\multicolumn{3}{c|}{Integrated luminosity} \\
\hline
$M_{Q}$  & 1 $\textnormal{fb}^{-1}$  & 5 $\textnormal{fb}^{-1}$  & 10 $\textnormal{fb}^{-1}$  \\ \hline
$400$               & 415     &  454       &   535 	   \\ \hline    
$600$               & 459     &  511       &   542 	    \\ \hline
$800$               & 459     &  515       &   544 	    \\ \hline   
\end{tabular}
\end{center}
\end{table}

\subsection*{sLHC}

The upgrade of the LHC machine, also referred as the sLHC \cite{SLHC} project aims at increasing the peak luminosity by 
a factor of 10 and deliver approximately 3000 $\textnormal{fb}^{-1}$ to the experiments. Although it's rather difficult to 
foresee what would be interesting to study at the sLHC without having the LHC run at its nominal luminosity first, 
here we assume that no vector bilepton signals were found at the LHC and explore the sLHC exclusion and discovery 
potentials for the exotic particles. Using the same techniques described in the previous sections we find the exclusion 
potential for three heavy quark masses: 1170 GeV for $M_{Q}$ = 400 GeV, 1220 GeV for $M_{Q}$ = 600 GeV and 1300 GeV for 
$M_{Q}$ = 800 GeV. The discovery potential is also increased: 1100 GeV for $M_{Q}$ = 400 GeV, 1150 GeV for $M_{Q}$ = 600 GeV 
and 1230 GeV for $M_{Q}$ = 800 GeV. This represents a gain of $\sim$ 200 GeV in terms of discovery mass reach compared to 
the default luminosity 14 TeV LHC run. This region is certainly worth exploring since it is still considerably 
below the upper limit of 3.5 TeV we have discussed in section II. The results for the sLHC are displayed in Table IV.

\begin{table}
  \begin{center}
    \caption[]{%
      Discovery mass reach and exclusion limits for doubly-charged bileptons with respect to sLHC and 331 exotic quark masses. Masses are in GeV.
    \label{tab:tab4}}
\begin{tabular}{|c|c|c|c|}
\hline
$M_{Q}$             & Discovery  & Exclusion    \\ \hline
400                 & 1100    &  1170       \\ \hline    
600                 & 1150    &  1220       \\ \hline
800                 & 1230    &  1300       \\ \hline   
\end{tabular}
\end{center}
\end{table}

\section{Conclusions}

We have investigated the LHC potential for discovering or setting limits on doubly-charged vector bileptons, in different
 scenarios considering important experimental aspects in the simulation like detector geometrical acceptance, 
luminosity uncertainty and lepton efficiency. By analysing the observable final state $e^{\mp}e^{\mp}\mu^{\pm}\mu^{\pm}$, 
we have found that bilepton signatures can already be observed at very early stages of LHC running with 14 TeV, 
if the bilepton mass is not much greater than 400 GeV. On the other hand, if the bilepton mass lies in the TeV scale, 
at least 10 years of the machine operation will be needed for discovering it, if $M_{Q} >$ 600 GeV. The observation 
of such signal, in combination with $Z^{\prime}_{331}$ searches in dilepton channel, would provide a very powerful way of 
discriminating between 331 models and other BSM scenarios which also predict heavy neutral gauge bosons.
If no signal is observed, the LHC can extend considerably the currents limits on bilepton mass by direct search in 
the four lepton final state. 

At the current LHC energy, 7 TeV, masses up to 566 GeV can be excluded and if 10 $\textnormal{fb}^{-1}$ of data is recorded,
vector bilepton masses up to 544 GeV could be discovered. We also found that the sLHC can expand the lower exclusion limits 
up to a mass of 1300 GeV. 

We also made a revision on current experimental bounds on bileptons in 331 models and the possibility of results 
from the LHC to exclude some versions of these models. We found that it is not possible to safely discard any 331 model, 
including its minimal version, at the LHC. Purely theoretical arguments taken from the literature are used to draw this 
conclusion.

Furthermore, we investigated how the heavy quark sector of the 331 model influences our results. In some cases a 
substantial change in the process cross-section is observed by varying the value of the heavy quark masses. 
Since some of the best previous limits on bileptons were coming from experiments containing at least one leptonic beam, 
we conclude that new results from the LHC will be indispensable in determining to a more accurate extent which models 
like 331 can be disfavored or discovered. The final state studied in this article will be the best channel 
to experimentally determine this.


\begin{acknowledgments}
This work has been partially supported by the European Commission within the 7th framework programme (FP7-PEOPLE-2009-IEF, Grant Agreement no. 253339). 
\end{acknowledgments}


\end{document}